\documentclass[preprint,aps,eqsecnum]{revtex4}
\begin{document}
%\draft
%\preprint{\vbox{\baselineskip=12pt}
%\rightline{}
%\rightline{hep-th/0512494}
%\title{Effective Einstein Equation : A new approach to astrophysical observations
%without dark matter}
\title{An alternative to dark matter:do braneworld effects hold the key?}
\author{Supratik Pal \footnote{Electronic address: 
{\em supratik@cts.iitkgp.ernet.in}}}
\affiliation{
Department of Physics and Centre for Theoretical Studies\\
Indian Institute of Technology\\ Kharagpur 721 302, INDIA\\ 
and\\
Relativity and Cosmology Research Centre\\ Department of Physics \\ 
Jadavpur University\\  Kolkata 700 032, INDIA}
\vspace{.5in}

\begin{abstract}

An alternative way of explaining astrophysical observations
without dark matter is proposed. In the braneworld scenario,
where our universe is visualised as a 4-dimensional hypersurface
embedded in a higher dimensional spacetime,  
the effective Einstein equation on the brane contains extra
terms. We show that the astrophysical
observations which are usually explained by particle dark 
matter, can be explained 
in this modified theory of gravity via those extra terms, without 
the need for dark matter. 
As specific example, we model X-ray
profiles of clusters of galaxies and the galaxy rotation curves which
are consistent with observations. Further, we investigate whether  
gravitational lensing can discriminate between the two possibilities.
Significant differences between the values of the deflection angles
are obtained, which can be tested in future observations.   

\end{abstract}

\maketitle

\section {Introduction : A look on dark matter}

An important issue of present-day astrophysics is the  
understanding of dark matter.  
These are apparently invisible objects but their existence is inevitable
in explaining several aspects of astrophysics and cosmology.
There are a number of possible candidates for dark matter. To
 mention a few, the particle dark matter models propose
`baryonic dark matter' which include
massive compact halo objects (MACHO). The second kind 
of these models are the `non-baryonic dark matter' which constitute a
pressureless medium.  
Examples include exotic supersymmetric particles, known as the weakly interacting 
massive particles, the WIMP {\cite {wimp1, wimp2}}.  Models with substantial
amount of pressure have also been proposed {\cite {sbsk}}.
These are all material candidates for dark matter. There are alternative
propositions, called the MOND {\cite {mond1, mond2}}, that
are based on the modifications of Newtonian dynamics. 
A relativistic version of the MOND is also available {\cite {bek}}.

Astrophysicists need to introduce dark matter in two contexts
{\cite {dm1, dm2}}. First, the observed X-ray emission from hot, intra-cluster gas 
can be used to measure its density and pressure. The mass estimated from these
data using Einstein's theory of gravitation
(the so-called Newtonian analysis) is always found to be substantially
in excess of what can be attributed to visible matter. So, to preserve
standard Einstein's theory, it is proposed that around $\sim 80 \%$ of the matter
surrounding clusters is dark.

Secondly, the Doppler shift measured from the  
neutral hydrogen cloud rotating inside spiral galaxies   
can be used to estimate the rotational velocity. It has been observed that
the rotational velocity decreases away from centre
but remains constant after reaching a typical value of $\sim 200$ km/s.
The velocity profile thus found, called the ``rotation curve'', when
applied to standard Einstein gravity,
gives rise to an enhanced mass. 
Hence if Einstein's theory has to be preserved, then one has to postulate that the  
galaxy is embedded in
a halo made up of some kind of invisible matter, called dark matter 
{\cite {curv1, curv2}}. 

However, one should note that Einstein's theory of gravitation has
not been tested in these length scales. 
%Hence it can be questioned : why should we take it as {\em the theory} ?  
Further, the exact nature of dark matter is not well-understood till date.
Hence one may start from the very beginning with a modified version of the
standard Einstein theory of gravity by which the above observations
might be explained without the need for dark matter.
Here is one such theory, popularly known as the effective Einstein equation.  
The term ``Effective Einstein Equation'' means that if there is an
($N - 1$)-dimensional hypersurface embedded in an $N$-dimensional spacetime
then the Einstein equation on
the hypersurface takes into account the effects of the terms coming
from the extra dimensions in the $N$-dimensional spacetime ($N > 4$).
Hence, the geometry and
physics on the hypersurface (our four dimensional world)
is governed by this  modified Einstein equation.
We propose that 
it is possible to model spiral galaxies and clusters using this effective 
Einstein equation, that can successfully replace dark matter with the 
effects coming from extra dimensions.

\section{Effective Einstein Equation : A review}

Let us now switch over to a brief review of the effective Einstein 
equation. Brane world physics {\cite {rs1, rs2}}
 predicts that the observable universe might be a 
4-dimensional hypersurface (3-brane)
embedded in a 5-dimensional (or higher) spacetime (bulk). 
A new geometrical approach to the  brane world physics, 
called the effective
Einstein equation {\cite {shiro}}, is in vogue for the last few years.
The basic notion is that if there is an extra dimension,
then it has to have some effect on our 4D universe and one
should take into account those effects in order to obtain
correct physics. 
This formalism helps us take a different look at the 
physical phenomena that are not well-established till date. Some such aspects 
have been studied in {\cite {ref1, ref2, padi, ref3, infla, mart, mak1, mak2}.

Without going into the technical details of how it is derived,
let us straightaway try to find out the essence of the effective 
Einstein equation on the brane which reads {\cite {shiro, padi}}

\begin{equation}
G_{\mu \nu} = -\Lambda g_{\mu \nu} + \kappa_4^2 T_{\mu \nu} +
\kappa_5^4 S_{\mu \nu} -E_{\mu \nu}
\label{eq:a1}
\end{equation}

Before proceeding further, let us have a look on the above equation. 
The first two terms in the expression of $G_{\mu \nu}$ combine to give
the standard 4D Einstein equation. The third term $S_{\mu \nu}$
is the quadratic contribution from the brane energy-momentum tensor
$T_{\mu \nu}$ and the last term $E_{\mu \nu}$
is the projected bulk Weyl tensor on the brane, in absence of bulk matter
{\cite{shiro, padi}}.
%\begin{equation}
%E_{MN} = C^I_{JKL} n_I n^K q_M^J q_N^L - \frac{2}{3} \left[ q_M^C q_N^D
%+n^C n^D q_{MN} - \frac{1}{4} g^{CD} q_{MN} \right] 8 \pi G T_{CD}
%\label{eq:a2}
%\end{equation}

The 4D cosmological constant $\Lambda$ can be set to zero by proper
fine-tuning {\cite {rs1, rs2}} and for weak field (which is relevant in
the present study) the contribution from $S_{\mu\nu}$ is negligibly
small {\cite {shiro}}. Hence Eq (\ref{eq:a1}) reduces to {\cite {ours,
proc}}

\begin{equation}
G_{\mu \nu} =  \kappa_4^2 T_{\mu \nu} - E_{\mu \nu}
\label{eq:a3}
\end{equation}

The net result is that the standard 4D Einstein equation is 
modified by a purely geometric term 
$E_{\mu \nu}$ which is a traceless symmetric tensor.
This is the property 
we are going to highlight in subsequent discussions.

\section{The scheme : Formalism without dark matter}

We are now in a position to replace dark matter by our formalism.
In particular, we look for modelling spiral galaxies
and clusters consistent with observed rotation curves or X-ray
profiles. Contrary to the 
usual concept, we assume that there is no such dark matter in form 
of halos surrounding the galaxies or the clusters but the visible matter is 
{\em all} that contributes.
Since the gravitational fields inside the galaxies and clusters are weak, 
the metric can be completely specified by two unknown
potentials: the Newtonian potential $\Phi(r)$ and the relativistic
one $\Psi(r)$, in isotropic coordinates as {\cite {ours}} : 

\begin{equation}
ds^2 = -(1 + 2 \Phi) dt^2 + (1 -2 \Phi + 2 \Psi) (dr^2 + r^2 d \theta^2 + r^2
\sin^2 \theta d \phi^2)
\label{eq:c1}
\end{equation}

The Einstein tensor components are listed below.
\begin{equation}
G^0_0=- 2 \nabla^2 (\Phi - \Psi) \,, \,  G^r_r=2 \frac{\Psi^{'}}{r} \,
, \,
G^{\theta}_{\theta}=G^{\phi}_{\phi}=\Psi^{''}+\frac{\Psi^{'}}{r}
\label{eq:c2}
\end{equation}

In the usual Newtonian analysis based on the standard Einstein equation
, and considering that $T_{00} = - \rho c^2$ is the only
non-zero component, $\Psi(r)$ comes out to be zero and we have
%\begin{equation}
%G^{\mu}_{\nu}=\frac{8 \pi G}{c^4} T^{\mu}_{\nu}
%\label{eq:a11}
%\end{equation} 
\begin{equation}
\nabla^2 \Phi = \frac{4 \pi G}{c^2} \rho
\label{eq:c3}
\end{equation} 
where $\Phi$ is determined directly from observations of either 
rotation curves (for spiral galaxies) or 
X-ray profiles (for galaxy clusters).  
The mass estimated from Eq ({\ref{eq:c3}}) with $\Phi(r)$ only
is {\em always} found to be substantially in excess of the visible matter.  
To solve the puzzle it was proposed that $\sim 80 \%$ 
of the matter surrounding the spiral galaxies and clusters
in form of halos is dark matter. 

Let us now look back to the essential information that emerges from the 
effective Einstein equation. Here we have an extra term $E_{\mu \nu}$
which is traceless.  
We exploit this property by taking trace of Eq (\ref{eq:a3})
to obtain {\cite {ours, proc}}
  
\begin{equation}
\nabla^2 (\Phi - 2 \Psi)= \frac{4 \pi G}{c^2 }\rho_v
\label{eq:c4}
\end{equation}

The message is now crystal clear. That we were {\em probably by
mistake} attributing to the Newtonian potential $\Phi$ is, in reality,
compensated by $\Psi$, the later being non-zero
in the present scenario.  
One must have noticed that here there is no use of dark matter since the visible
matter with density $\rho_v$ is all that contributes. 

Solving Eq (\ref{eq:c4}) we arrive at {\cite {ours}} 
\begin{equation}
\Psi=\frac{1}{2} \Phi - \frac{2 \pi G}{c^2} (\nabla^2)^{-1} \rho_v \, 
\label{eq:c5}
\end{equation}

%The tensor $E^{\mu}_{\nu}$ can be calculated
%using eq. (\ref{eq:a2}) once both $\Phi$ and $\Psi$ are known. 

The bottomline is that we have a solution for the spiral galaxies
and clusters, consistent with observations, {\em without} the need for 
dark matter.

\subsection{Modelling clusters with X-ray emission}

Let us now apply the formalism to model galaxy clusters.
The observed X-ray emission from hot intra-cluster gas 
allow us measure its
density profile $\rho_g(r)$.
%and the pressure $P_g(r)= (k T/\mu m_p)\,  \rho_g(r)$.
The model that best fits this density profile is 
the isothermal $\beta$ model which reads {\cite {ours}}
\begin{equation}
\rho_g(r)=\rho_0 [1 + (r/r_c)^2]^{-3 \beta/2}
\label{eq:d2} 
\end{equation}

%Considering the energy-momentum tensor
%for the intra-cluster gas $T[{\rm gas}]_{\mu \nu}=(P_g + \rho_g c^2)
%U_{\mu} U_{\nu} - P_g g_{\mu \nu}$ and justifying that $P_g \ll \rho_g c^2$,
%projected energy-momentum  conservation 
%${T[{\rm  gas}]^{\mu}}_{\nu;\mu} =0$ orthogonal to $U_{\mu}$ gives
%{\cite {mtw}}
Use of the energy-momentum  conservation leads to the following equation
\begin{equation}
\Phi^{'}(r)=-\frac{ kT}{\mu m_p c^2}  \frac{ d \ln \rho_g}{d
  r} \,
\label{eq:d1} 
\end{equation}

Assuming  $\beta=2/3$ and restricting our analysis to  $r \gg r_c$
we obtain the Newtonian potential   
\begin{equation}
\Phi(r) = \frac{2kT}{\mu m_p c^2} \ln \frac{r}{r_c} \,
\label{eq:d3}
\end{equation}

Since there is no dark matter, {\em ie}, $\rho_v=\rho_g$ one can substitute
the expressions for $\rho_g$ and $\Phi$ into Eq ({\ref{eq:c4}}) to obtain 
{\cite {ours, proc}}

\begin{equation}
\Psi(r) = \left [ \frac{kT}{\mu m_p c^2}-\frac{2\pi G\rho_o r_c^2}{c^2} \right ]
\ln \frac{r}{r_c} 
\label{eq:d4}
\end{equation}
that compensates for the extra gravitational acceleration. 

%The non-zero components of  $E^{\mu}_{\nu}$,  $E^0_0=-E^r_r=[(
%  kT/\mu m_p c^2)-(4 \pi G \rho_0 r_c^2/c^2)] r^{-2}$ now  
%   provide the extra gravitational acceleration. 

\subsection{Explaining rotation curves of spiral galaxies}

Inside a spiral galaxy visible matter is mainly distributed in a disk the 
density profile of which is modelled as {\cite {mann}}

\begin{equation}
\rho_v = \rho_0 e^{- \gamma R^{'}} \delta (z^{'})
\label{eq:e1}
\end{equation}
where $z = z^{'}$ is the position of the disk and $\gamma = {1 \over r_c}$,
the inverse scale length for the disk. 

The circular velocity  $v_c(r)$, referred to as ``rotation curve",  
can be used to obtain the geodesic equation of the 
neutral hydrogen (HI) clouds rotating inside spiral galaxies. 
\begin{equation}
\Phi^{'}(r) = \frac{v_c^2(r)}{c^2} {1 \over r}
\label{eq:e2}
\end{equation}

The Newtonian potential $\Phi(r)$  can be readily obtained from the
above equation {\cite {sbsk}}

\begin{equation}
\Phi(r) =\frac{v_c^2}{c^2} \left[\ln \left(\frac {r}{r_0} \right)-1 \right]
\label{eq:e3}
\end{equation}
Note that the solution matches the exterior
Scwarzshild metric at the boundary $r =r_0$.

Use of the density profile in Eq ({\ref{eq:c4}})
gives the Green function solution for $\Phi - 2 \Psi$, 
which reads for large $R$ 

\begin{equation}
\Phi- 2 \Psi = \left[ \frac{8 \pi^2 G \rho_0}{\gamma^2 c^2 } \right]
\frac{1}{\sqrt{R^2 + |z|^2}}
\label{eq:e4}
\end{equation}
wherefrom it is a trivial job to find out $\Psi$ as

\begin{equation}
\Psi(r) = \frac{v_c^2}{2 c^2} \left[\ln \left(\frac {r}{r_0} \right)-1 \right]
- \left[\frac{4 \pi^2 G \rho_0}{\gamma^2 c^2}\right] {1 \over r} 
\label{eq:e5}
\end{equation}

Hence far from the origin, the potentials look spherically symmetric, which reveals
that the disk appears as a point source to a very distant observer.

\section{Gravitational lensing : Alternative probe}

So far we have learned that the extra dimensional scenario can successfully replace
dark matter by geometry. To be confirmed, one has to provide other observational
techniques that can discriminate between the two scenarios. Here we suggest that 
gravitational lensing data can be an alternative probe that can possibly
help us determine which scenario is correct.

In the dark matter concept, the deflection angle $\hat \alpha_N$ of a 
photon passing 
through the halo is fixed by the Newtonian potential $\Phi$ only. 

\begin{equation}
\hat{\alpha}_N = 2 \int_{s}^{o}  \hat{\nabla}_\perp  \Phi
\, \,  dl \ 
\label{eq:f1}
\end{equation}

On the other hand, in the effective Einstein equation scenario where $\Psi \neq 0$, 
the deflection angle is modified to {\cite {ours}}

\begin{equation}
\hat{\alpha } =  \int_{s}^{o} \hat{\nabla}_\perp (2 \Phi -
\Psi)  \, \,  dl  \
\label{eq:f2}
\end{equation}

Let us see by how much amount the modified deflection angle differs
from its Newtonian value.

\subsection{For clusters}

For our choice of clusters, the Newtonian deflection angle is found
to be
\begin{equation}
\hat{\alpha_N}=\frac{4\pi kT}{\mu m_p c^2}
\label{eq:f3}
\end{equation}
whereas the deflection angle from the proposed scenario turns out to be
%{\cite {ours, lens}}
\begin{equation}
\hat{\alpha} = \hat{\alpha_N} \left[ 0.75 + \frac{\pi G \rho_0
r_c^2 \mu m_p}{2 k T}\right]  
\label{eq:f4}
\end{equation}

\subsection{For spiral galaxies}

In case of spiral galaxies, the Newtonian deflection angle is the usual one
\begin{equation}
\hat{\alpha_N}= \frac{2 \pi v_c^2}{c^2}
\label{eq:f5}
\end{equation}
and the modified deflection angle is given by 
%{\cite {disk, lens}}
\begin{equation}
\hat{\alpha } = \hat{\alpha_N} \left[ 0.75 - \frac{4 \pi G \rho_0}
{\gamma^2 v_c^2 b}\right]  
\label{eq:f6}
\end{equation}

The calculations above reveal that
the deflection angle in the present scenario is around $\sim 75 \%$ of
the Newtonian value. Unfortunately, this difference falls within the present
experimental error bars. We look forward for sufficiently accurate lensing 
data for a more conclusive remark.

\section{Summary and outlook}

Here we have proposed an alternative to dark matter in astrophysics. 
We note the standard Einstein theory of gravity
has not been tested on galactic length scales, yet it has been applied
to estimate the mass of galaxies and clusters, that give rise to an
inevitable prediction of dark matter. In stead of using the standard
Einstein equation, we have applied its modified version, namely, the
effective Einstein equation to explain the observed
rotation curves of spiral galaxies and the X-ray emission from galaxy
clusters. Thus, even in absence of 
dark matter, these astrophysical observations can possibly be realised as the 
effect of the embedding geometry alone. 

We have also succeeded in providing an alternative observational technique
that can discriminate between the two scenarios. This technique, namely,
gravitational lensing calculation, reveals that the proposed scenario
differs by a factor of $\sim 25 \%$ from the dark matter analysis, that
can be subjected to observational verification in future.
  
To conclude, our basic goal was to explain astrophysical observations without 
dark matter.  But the context of dark
matter also arises in cosmology to explain the expanding universe. 
Even models of galaxies with relativistic stresses are around {\cite {lete}}.
It remains as further work to see if these aspects of gravity can be explained 
by the extra dimensional effects.

\section*{Acknowledgment :} 
The author thanks Sayan Kar and Somnath Bharadwaj for collaborations
and valuable suggestions.

\end{document}